\documentclass[aps,prd,onecolumn,groupedaddress,showpacs,nofootinbib,amssymb]{revtex4}
\usepackage[dvips]{graphicx}
\usepackage{amssymb}
\usepackage{amsmath}
\usepackage{graphicx,,color}
\usepackage{amsfonts}
\usepackage{bm}
\usepackage{cancel}
\usepackage{comment}

\newcommand\be{\begin{equation}}
\newcommand\ee{\end{equation}}

\allowdisplaybreaks[4]

\begin{document}

\tolerance=5000

\title{Reconstructing ACT-compatible and GW170817-compatible Einstein-Gauss-Bonnet Inflation from the Observational Indices}
\author{V.K. Oikonomou$^{1,2}$}\email{voikonomou@gapps.auth.gr;v.k.oikonomou1979@gmail.com}
\affiliation{$^{1)}$Department of Physics, Aristotle University of
Thessaloniki, Thessaloniki 54124, Greece\\
$^{2)}$ Center for Theoretical Physics, Khazar University, 41
Mehseti Str., Baku, AZ-1096, Azerbaijan}

\tolerance=5000

\begin{abstract}
In this work we use an inverse reconstruction technique for
constructing ACT-compatible and GW170817-compatible
Einstein-Gauss-Bonnet inflationary theories. From a given
tensor-to-scalar ratio using the reconstruction technique, we find
which scalar Gauss-Bonnet coupling function and which scalar
potential can yield the given tensor-to-scalar ratio. We present
the formalism and the viable theories pass a series of
observational tests, including the amplitude of the scalar
perturbations, which is non-trivial for Einstein-Gauss-Bonnet
theories. We present four viable models of inflation that pass all
the observational tests.
\end{abstract}


\maketitle

\section{Introduction}

The last decade was particularly fruitful from the point of view
of observational astrophysics, since major events were reported.
The chorus of remarkable observations started with the GW170817
event back in 2017
\cite{TheLIGOScientific:2017qsa,Monitor:2017mdv,GBM:2017lvd},
which utterly changed the way of thinking in theoretical
cosmology. Specifically, in the GW170817 event, the gravitational
wave and the electromagnetic signal observed by the kilonova
arrived almost simultaneously, a fact that severely constrained
the gravitational wave speed to be $|c_T/c-1|\leq 5 \times
10^{-16}$. This alone immediately excluded the massive graviton
theories, theories that predict tensor perturbations with
propagation speed distinct from that of light
\cite{Ezquiaga:2017ekz,Baker:2017hug,Creminelli:2017sry,Sakstein:2017xjx,Fernandes:2022zrq}.
One of these theories was Einstein-Gauss-Bonnet theory
\cite{Hwang:2005hb,Nojiri:2006je,Cognola:2006sp,Nojiri:2005vv,Nojiri:2005jg,Satoh:2007gn,
Bamba:2014zoa,Yi:2018gse,Guo:2009uk,Guo:2010jr,Jiang:2013gza,vandeBruck:2017voa,Pozdeeva:2020apf,
Vernov:2021hxo,Pozdeeva:2021iwc,Fomin:2020hfh,DeLaurentis:2015fea,Chervon:2019sey,Nozari:2017rta,
Odintsov:2018zhw,Kawai:1998ab,Yi:2018dhl,vandeBruck:2016xvt,Maeda:2011zn,Ai:2020peo,Easther:1996yd,Codello:2015mba,
Oikonomou:2021kql,Odintsov:2020sqy,Oikonomou:2024etl,Fier:2025huc,Oikonomou:2022ksx,Oikonomou:2022xoq},
which however was revived in a series of articles that reformed
their inflationary framework
\cite{Odintsov:2020sqy,Oikonomou:2021kql}, see also Ref.
\cite{Oikonomou:2024etl} for a recent framework in the same line
of research. These theories are vital for inflation since they
lead sometimes to a positive tensor spectral index, a feature
absent in ordinary scalar field inflation.

On the other hand, the inflationary regime
\cite{inflation1,inflation2,inflation3,inflation4} will be
thoroughly tested in the next ten years, both at the cosmic
microwave background (CMB) level
\cite{SimonsObservatory:2019qwx,LiteBIRD:2022cnt} and indirectly
by the gravitational wave experiments
\cite{Hild:2010id,Baker:2019nia,Smith:2019wny,Crowder:2005nr,Smith:2016jqs,Seto:2001qf,Kawamura:2020pcg,Bull:2018lat,LISACosmologyWorkingGroup:2022jok}.
The existence of a stochastic gravitational wave background may be
linked to the inflationary era, and Einstein-Gauss-Bonnet theories
with a positive tensor spectral index can generate detectable
stochastic gravitational wave signals \cite{Oikonomou:2022xoq}.
Such a stochastic gravitational wave background has been detected
in 2023 by various pulsar timing array experiments like NANOGrav
\cite{NANOGrav:2023gor}, but inflation does not suffice in
explaining the signal \cite{Vagnozzi:2023lwo,Oikonomou:2023qfz}.
One of the candidate theories towards explaining a stochastic
gravitational wave signal by LISA, is Einstein-Gauss-Bonnet
theory, which is a prominent modified gravity theory
\cite{reviews1,reviews2,reviews3,reviews4}. Now, recently, the
plot thickened with inflation, since the ACT data in 2025
indicated that the spectral index of the scalar perturbations is
2$\sigma$ off the Planck value, and specifically it is constrained
as follows \cite{ACT:2025fju, ACT:2025tim},
\begin{align}
\label{act} n_{s}=0.9743 \pm 0.0034\, .
\end{align}
If we combine this with the updated Planck/BICEP constraints on
the tensor-to-scalar ratio \cite{BICEP:2021xfz},
\begin{align}
\label{planck} r<0.036\, ,
\end{align}
the above two results severely constrained the inflationary models
that can be viable. Even traditional inflationary models were
rendered non-viable in one night. After the ACT data were
released, a large stream of articles that aimed to reconcile
inflationary theories with the ACT data appeared, see Refs.
\cite{Kallosh:2025rni,Gao:2025onc,Liu:2025qca,Yogesh:2025wak,Yi:2025dms,Peng:2025bws,Yin:2025rrs,Byrnes:2025kit,
Wolf:2025ecy,Aoki:2025wld,Gao:2025viy,Zahoor:2025nuq,Ferreira:2025lrd,Mohammadi:2025gbu,Choudhury:2025vso,
Odintsov:2025wai,Q:2025ycf,Zhu:2025twm,Kouniatalis:2025orn,Hai:2025wvs,Dioguardi:2025vci,Yuennan:2025kde,
Kuralkar:2025zxr,Kuralkar:2025hoz,Modak:2025bjv,Oikonomou:2025xms,Odintsov:2025jky,
Aoki:2025ywt,Ahghari:2025hfy,McDonough:2025lzo,Chakraborty:2025wqn,NooriGashti:2025gug,Yuennan:2025mlg,
Deb:2025gtk,Afshar:2025ndm,Ellis:2025zrf,Yuennan:2025tyx,Wang:2025cpp,Qiu:2025uot,Wang:2025dbj,Asaka:2015vza,Oikonomou:2025htz,Choudhury:2025hnu,Singh:2025uyr,Kim:2025dyi}.
In addition, in Ref. \cite{Peng:2026ofs} an artifical intelligence
program produced ACT-compatible inflation theories.

In view of the above line of research, in this work we aim to
provide a new reconstruction scheme in Einstein-Gauss-Bonnet
theories, in which we start from the observational indices and
specifically the tensor-to-scalar ratio, and work our way up
building the model that can yield such a tensor-to-scalar ratio.
In this way we construct several viable models of
Einstein-Gauss-Bonnet gravity which can be compatible with both
the ACT data and the updated BICEP/Planck data. The viability of
the Einstein-Gauss-Bonnet theory is not a simple task, because
apart from the observational indices, one must succeed in having
an amplitude of the scalar perturbations which must also be
compatible with the Planck constraints. All the studied models
satisfy a series of constraints and thus are viable
phenomenologically.

\section{Overview of GW170817-Compatible Einstein-Gauss-Bonnet Inflation and Reconstruction from the Indices}

Let us briefly review the GW170817-compatible formalism of
Einstein-Gauss-Bonnet inflation, based on Ref.
\cite{Oikonomou:2021kql}. Then we shall demonstrate how to obtain
viable Einstein-Gauss-Bonnet inflation by starting from the
observational indices. The Einstein-Gauss-Bonnet gravity action in
vacuum is,
\begin{equation}
\label{action} \centering
S=\int{d^4x\sqrt{-g}\left(\frac{R}{2\kappa^2}-\frac{1}{2}\partial_{\mu}\phi\partial^{\mu}\phi-V(\phi)-\frac{1}{2}\xi(\phi)\mathcal{G}\right)}\,
,
\end{equation}
where $R$ denotes the Ricci scalar, and $\kappa=\frac{1}{M_p}$
where $M_p$ is the reduced Planck mass. In addition, $\mathcal{G}$
stands for the Gauss-Bonnet invariant in four dimension, which is
defined as
$\mathcal{G}=R^2-4R_{\alpha\beta}R^{\alpha\beta}+R_{\alpha\beta\gamma\delta}R^{\alpha\beta\gamma\delta}$
with $R_{\alpha\beta}$ and $R_{\alpha\beta\gamma\delta}$ being the
Ricci and Riemann tensor respectively. We shall assume that the
metric is a flat Friedmann-Robertson-Walker (FRW) metric, with
line element being,
\begin{equation}
\label{metric} \centering
ds^2=-dt^2+a(t)^2\sum_{i=1}^{3}{(dx^{i})^2}\, ,
\end{equation}
and with $a$ being the scale factor. For the FRW metric, the
Gauss-Bonnet invariant is $\mathcal{G}=24H^2(\dot H+H^2)$. By
assuming that the scalar field is only time-dependent, by varying
the action with respect to the metric and the scalar field, we
obtain the field equations,
\begin{equation}
\label{motion1} \centering
\frac{3H^2}{\kappa^2}=\frac{1}{2}\dot\phi^2+V+12 \dot\xi H^3\, ,
\end{equation}
\begin{equation}
\label{motion2} \centering \frac{2\dot
H}{\kappa^2}=-\dot\phi^2+4\ddot\xi H^2+8\dot\xi H\dot H-4\dot\xi
H^3\, ,
\end{equation}
\begin{equation}
\label{motion3} \centering \ddot\phi+3H\dot\phi+V'+12 \xi'H^2(\dot
H+H^2)=0\, ,
\end{equation}
with the ``dot'' denoting differentiations with respect to the
cosmic time $t$ and the ``prime'' denotes differentiation with
respect to the scalar field. We impose the following slow-roll
conditions,
\begin{equation}\label{slowrollhubble}
\dot{H}\ll H^2,\,\,\ \frac{\dot\phi^2}{2} \ll V,\,\,\,\ddot\phi\ll
3 H\dot\phi\, .
\end{equation}
and also we impose the constraint $c_T^2=1$, with $c_T^2$ being
\cite{Hwang:2005hb},
\begin{equation}
\label{GW} \centering c_T^2=1-\frac{Q_f}{2Q_t}\, ,
\end{equation}
and also $Q_f$ and $Q_b$ appearing above being $Q_f=8
(\ddot\xi-H\dot\xi)$, $Q_t=F+\frac{Q_b}{2}$, with $F$ being equal
to $F=\frac{1}{\kappa^2}$ while $Q_b=-8 \dot\xi H$. In view of the
above, the simplified field equations become,
\begin{equation}
\label{motion5} \centering H^2\simeq\frac{\kappa^2V}{3}\, ,
\end{equation}
\begin{equation}
\label{motion6} \centering \dot H\simeq-\frac{1}{2}\kappa^2
\dot\phi^2\, ,
\end{equation}
\begin{equation}
\label{motion8} \centering \dot\phi\simeq\frac{H\xi'}{\xi''}\, .
\end{equation}
In addition, the scalar field potential and the Gauss-Bonnet
scalar coupling function $\xi(\phi)$ must satisfy the following
differential equation,
\begin{equation}
\label{maindiffeqnnew} \centering
\frac{V'}{V^2}+\frac{4\kappa^4}{3}\xi'\simeq 0\, ,
\end{equation}
which constraints their functional form, thus these are not free
to choose in an independent way. For Einstein-Gauss-Bonnet models
of inflation, the slow-roll indices are defined as follows
\cite{Hwang:2005hb},
\begin{align}\label{slowrollbasic}
& \epsilon_1=-\frac{\dot
H}{H^2},\,\,\,\epsilon_2=\frac{\ddot\phi}{H\dot\phi},
\,\,\,\epsilon_4=\frac{\dot E}{2HE},
\\ \notag &
\epsilon_5=\frac{Q_a}{2HQ_t},\,\,\, \epsilon_6=\frac{\dot
Q_t}{2HQ_t}\, ,
\end{align}
where $E$ is,
\begin{equation}\label{functionE}
E=\frac{F}{\dot\phi^2}\left(
\dot{\phi}^2+3\left(\frac{Q_a^2}{2Q_t}\right)\right)\, ,
\end{equation}
and $Q_a$ is defined as \cite{Hwang:2005hb},
\begin{align}\label{qis}
& Q_a=-4 \dot\xi H^2\, .
\end{align}
By using Eqs. (\ref{motion5})-(\ref{motion8}), the slow-roll
indices take the following final form,
\begin{equation}
\label{index1} \centering \epsilon_1\simeq\frac{\kappa^2
}{2}\left(\frac{\xi'}{\xi''}\right)^2\, ,
\end{equation}
\begin{equation}
\label{index2} \centering
\epsilon_2\simeq1-\epsilon_1-\frac{\xi'\xi'''}{\xi''^2}\, ,
\end{equation}
\begin{equation}
\label{index4} \centering
\epsilon_4\simeq\frac{\xi'}{2\xi''}\frac{\mathcal{E}'}{\mathcal{E}}\,
,
\end{equation}
\begin{equation}
\label{index5} \centering
\epsilon_5\simeq-\frac{\epsilon_1}{\lambda}\, ,
\end{equation}
\begin{equation}
\label{index6} \centering \epsilon_6\simeq
\epsilon_5(1-\epsilon_1)\, ,
\end{equation}
where, $\mathcal{E}=\mathcal{E}(\phi)$ and $\lambda=\lambda(\phi)$
is defined in the following way,
\begin{equation}\label{functionE}
\mathcal{E}(\phi)=\frac{1}{\kappa^2}\left(
1+72\frac{\epsilon_1^2}{\lambda^2} \right),\,\, \,
\lambda(\phi)=\frac{3}{4\xi''\kappa^2 V}\, .
\end{equation}
The definition of the observational indices in terms of the
slow-roll indices is \cite{Hwang:2005hb},
\begin{equation}
\label{spectralindex} \centering
n_{\mathcal{S}}=1-4\epsilon_1-2\epsilon_2-2\epsilon_4\, ,
\end{equation}
regarding the spectral index of the scalar primordial
perturbations, and regarding the tensor spectral index is
\cite{Oikonomou:2021kql},
\begin{equation}\label{tensorspectralindexfinal}
n_{\mathcal{T}}\simeq -2\epsilon_1\left ( 1-\frac{1}{\lambda
}+\frac{\epsilon_1}{\lambda}\right)\, .
\end{equation}
In addition, the final form of the tensor-to-scalar ratio is
remarkably simplified \cite{Oikonomou:2021kql},
\begin{equation}\label{tensortoscalarratiofinal}
r\simeq 16\epsilon_1\, .
\end{equation}
In addition, a viable inflationary theory must have a correct
amplitude of scalar perturbations which must be compatible with
the latest Planck data CMB constraint \cite{Planck:2018jri}
$\mathcal{P}_{\zeta}(k_*)=2.196^{+0.051}_{-0.06}\times 10^{-9}$.
The amplitude of the scalar perturbations for
Einstein-Gauss-Bonnet theories, in terms of the slow-roll indices,
is defined to be \cite{Hwang:2005hb},
\begin{equation}\label{powerspectrumscalaramplitude}
\mathcal{P}_{\zeta}(k)=\left(\frac{k \left((-2
\epsilon_1-\epsilon_2-\epsilon_4) \left(0.57\, +\log \left(\left|k
\eta_* \right| \right)-2+\log (2)\right)-\epsilon_1+1\right)}{(2
\pi ) \left(z c_A^{\frac{4-n_{\mathcal{S}}}{2}}\right)}\right)^2\,
,
\end{equation}
where $z=\frac{a \dot{\phi} \sqrt{\frac{E(\phi )}{\frac{1}{\kappa
^2}}}}{H (\epsilon_5+1)}$ and
$\eta_*=-\frac{1}{aH}\frac{1}{-\epsilon_1+1}$, with $\eta_*$ being
the conformal time at horizon crossing which must be evaluated
when the first horizon crossing occurs, that is when
$c_A\,k=a\,H$.

Now let us utilize the formalism above, and let us briefly present
the reconstruction of viable Einstein-Gauss-Bonnet theories
directly from the observational indices, and specifically from the
tensor-to-scalar ratio. We must express every function, as a
function of the $e$-foldings number, and this is achieved if we
use,
\begin{equation}
\label{xiofN} \centering \xi'(\phi) = \frac{d N}{d
\phi}\frac{d\xi}{dN} ,
\end{equation}
and also,
\begin{equation}
\xi^{\prime \prime}=\frac{d \xi^{\prime}}{d \phi}=\frac{d N}{d
\phi} \frac{d}{d N}\left(\frac{d N}{d \phi} \frac{d \xi}{d
N}\right)=\left(\frac{d N}{d \phi}\right)^{2} \frac{d^{2} \xi}{d
N^{2}}+\xi^{\prime} \frac{d}{d N}\left(\frac{d N}{d \phi}\right)
,
\end{equation}
Using Eqs. (\ref{index1}), (\ref{tensortoscalarratiofinal}) and
also the following,
\begin{equation}
\label{xiprimeprimeoverxiprime} \centering
\frac{\xi''}{\xi'}=\frac{d N}{d \phi}
,
\end{equation}
we get,
\begin{equation}
\label{finaltensortoscalar} \centering r(N)=8\kappa^2
\left(\frac{d \phi}{d N}\right)^2 ,
\end{equation}
which is basically, the general functional form of the
tensor-to-scalar ratio in terms of the $e$-foldings number. The
above, can be written as,
\begin{equation}
\label{Nprime} \centering
\frac{d N}{d\phi}=\frac{2 \kappa \sqrt{2}}{\sqrt{r}}
,
\end{equation}
and therefore we can obtain a useful expression for the functions
$\xi'$ and $\xi''$, as follows,
\begin{equation}
\centering
\xi'=\frac{d N}{d \phi}\frac{d \xi}{d N}=\frac{2\kappa\sqrt{2}}{\sqrt{r}}\frac{d \xi}{d N}
,
\end{equation}
\begin{equation}
\xi^{\prime \prime}=\frac{8 \kappa^{2}}{r} \frac{d^{2} \xi}{d N^{2}}-\frac{4 \kappa^{2}}{r^{2}} \frac{d r}{d N} \frac{d \xi}{d N}
.
\end{equation}
Using the  above, and also the fact that $\xi''=\frac{d N}{d
\phi}\xi'$, we get the following,
\begin{equation}
\label{xiprimeprimefinal} \xi'' = \frac{8 \kappa^{2}}{r}
\frac{d^{2} \xi}{d N^{2}}-\frac{4 \kappa^{2}}{r^{2}} \frac{d r}{d
N} \frac{d \xi}{d N}=\frac{2 \kappa \sqrt{2}}{\sqrt{r}} \frac{2
\kappa \sqrt{2}}{\sqrt{r}} \frac{d \xi}{d N}=\frac{8
\kappa^{2}}{r} \frac{d \xi}{d N} .
\end{equation}
From Eq. (\ref{xiprimeprimefinal}) we can derive a differential
equation obeyed by the Gauss-Bonnet coupling function $\xi(N)$,
\begin{equation}
\label{difxi} \centering \frac{d^2 \xi}{d
N^2}-\left(\frac{1}{2r}\frac{d r}{d N}+1\right)\frac{d \xi}{d N}=0
.
\end{equation}
From Eq. (\ref{maindiffeqnnew}), we can see that the potential of
the scalar field has the following form,
\begin{equation}
\label{Vwithphi} \frac{1}{V^{2}} \frac{d V}{d \phi}+\frac{4
\kappa^{4}}{3} \frac{d \xi}{d \phi}=0 ,
\end{equation}
thus by using Eqs. (\ref{Vwithphi}), (\ref{xiofN}) we obtain,
\begin{equation}
\label{VwithN} \frac{1}{V^{2}} \frac{d N}{d \phi} \frac{d V}{d
N}+\frac{4 \kappa^{4}}{3} \frac{d N}{d \phi} \frac{d \xi}{d N}=0 ,
\end{equation}
which has the following solution,
\begin{equation}
\label{genpot} \centering
V(N)=\frac{3}{4\kappa^4}\frac{1}{\xi(N)}\, .
\end{equation}
This reconstruction technique from the tensor-to-scalar ratio
given, enables us to find the Gauss-Bonnet scalar coupling
function, the scalar field potential and the spectral indices
$n_\mathcal{S}$, $n_T$ from the given functional form of the
tensor-to-scalar ratio, as functions of the $e$-foldings number.
The technique is highly restricted, because the observational
indices must be compatible with the ACT data and the updated
BICEP/Planck constrains on the tensor-to-scalar ratio, but also
the amplitude of the scalar perturbations must be compatible with
the Planck CMB constraints. Using this method, in the next section
we present some characteristic examples of viable ACT-compatible
inflationary scenarios, reconstructed from the tensor-to-scalar
ratio.

\section{Examples of ACT-compatible Reconstructed Einstein-Gauss-Bonnet Inflation from the Indices}

Some simple examples of viable inflation can be obtained by
choosing $r=\delta/N^d$ where $d>0$ and $d=$even, and also from
the functional form $r=\delta e^{-\beta N}$, which are quite
simple forms. Let us analyze these in detail.

\subsection{Model I: Tensor-to-scalar ratio of the Form $r=\delta/N^2$}

Let us start with the case,
\begin{equation}
\label{r1} \centering r=\frac{\delta}{N^2} .
\end{equation}
\begin{figure}[h!]
\centering
\includegraphics[width=20pc]{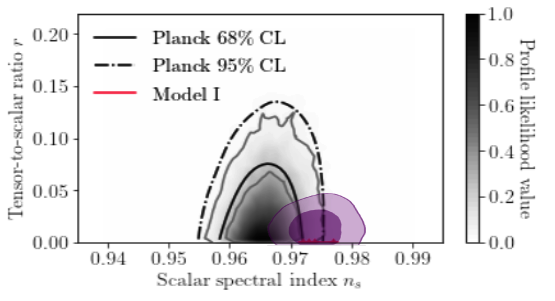}
\caption{Planck 2018 Likelihood Curves and ACT data for the Model
I with $r=\delta/N^2$ model}\label{plot1}
\end{figure}
From Eq. (\ref{difxi}) the scalar coupling function $\xi(N)$ is
obtained,
\begin{equation}
\label{xi1} \centering \xi(N)=C_1 \text{Ei}(N)+C_2 ,
\end{equation}
where $\text{Ei}(N)$ is the exponential integral function and
$C_1$, $C_2$ are dimensionless integration constants. The
corresponding potential can be found from (\ref{genpot}) and it
is,
\begin{equation}
\label{V1} \centering V(N)=\frac{3}{4 \kappa ^4 (C_1
\text{Ei}(N)+C_2)} \, .
\end{equation}
Accordingly, the function $\lambda(N)$ (\ref{functionE}) is,
\begin{equation}
\label{} \centering \lambda(N)=\frac{\delta  e^{-N} (C_1
\text{Ei}(N)+C_2)}{8 C_1 N} \, ,
\end{equation}
and by using Eq. (\ref{tensorspectralindexfinal}) the tensor
spectral index becomes,
\begin{equation}
\label{nT1} \centering n_T=-\frac{\delta  \left(1-\frac{C_1
e^{N}}{2 C_1 N \text{Ei}(N)+2 C_2 N}\right)}{8 N^2} .
\end{equation}
The spectral index of the scalar perturbations is too complicated
to quote it here, so we refrain from quoting it here. The model at
hand is easily viable for a wide range of the parameter values.
For example, for $\delta=0.0747$, $N=60$ and by choosing the free
parameters $C_1$ and $C_1$ as follows,
\begin{equation}\label{model1}
C_2=-1.93619\times 10^{24} C_1\, ,
\end{equation}
for $C_1=2.4\times 10^{-11}$, we obtain $n_{\mathcal{S}}=0.97754$,
$r=0.00025277$, $n_T=-0.50833$ and finally the predicted amplitude
of the scalar perturbations is
$\mathcal{P}_{\zeta}(k_*)=2.1543\times 10^{-9}$, thus the model is
Planck and ACT-compatible. In Fig. \ref{plot1} we present the
confrontation of the model with Planck 2018 likelihood curves and
the ACT data for $\delta=[0.0747, 0.91]$. As it can be seen, the
model is well fitted within the data.

\subsection{Model II: Tensor-to-scalar ratio of the Form $r=\delta/N^4$}

We continue with the case,
\begin{equation}
\label{r12} \centering r=\frac{\delta}{N^4} .
\end{equation}
\begin{figure}[h!]
\centering
\includegraphics[width=20pc]{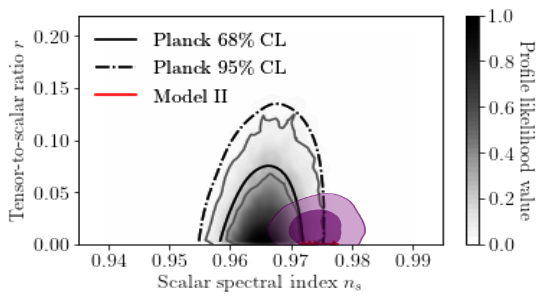}
\caption{Planck 2018 Likelihood Curves and ACT data for the Model
II with $r=\delta/N^4$ model}\label{plot2}
\end{figure}
From Eq. (\ref{difxi}) the scalar coupling function $\xi(N)$ is
obtained and in this case we have,
\begin{equation}
\label{xi12} \centering \xi(N)=C_1
\left(\text{Ei}(N)-\frac{e^{N}}{N}\right)+C_2 ,
\end{equation}
where $\text{Ei}(N)$ being again the exponential integral function
and $C_1$, $C_2$ are again dimensionless integration constants.
The corresponding potential can be found from (\ref{genpot}) and
it is,
\begin{equation}
\label{V12} \centering V(N)=\frac{3}{4 \kappa ^4 \left(C_1
\left(\text{Ei}(N)-\frac{e^{N}}{N}\right)+C_2\right)} \, .
\end{equation}
Accordingly, the function $\lambda(N)$ (\ref{functionE}) is in
this case,
\begin{equation}
\label{} \centering \lambda(N)=\frac{\delta  e^{-N} \left(C_1
\left(\text{Ei}(N)-\frac{e^{N}}{N}\right)+C_2\right)}{8 C_1 N^2}
\, ,
\end{equation}
and by using Eq. (\ref{tensorspectralindexfinal}) the tensor
spectral index becomes in this case,
\begin{equation}
\label{nT12} \centering n_T=-\frac{\delta  \left(\frac{C_1
e^{N}}{-2 C_1 N^2 \text{Ei}(N)+2 C_1 e^{N} N-2 C_2
N^2}+1\right)}{8 N^4} .
\end{equation}
In this case too, the spectral index of the scalar perturbations
is too complicated to quote it here, so we refrain from quoting it
here. The model at hand is easily viable for a wide range of the
parameter values. For example, for $\delta=3003$, $N=60$ and by
choosing the free parameters $C_1$ and $C_1$ as follows,
\begin{equation}\label{model12}
C_2=-3.28375\times 10^{22} C_1\, ,
\end{equation}
for $C_1=0.65\times 10^{-4}$, we obtain $n_{\mathcal{S}}=0.9740$,
$r=0.00046373$, $n_T=-0.5166$ and finally the predicted amplitude
of the scalar perturbations is
$\mathcal{P}_{\zeta}(k_*)=2.1526\times 10^{-9}$, thus the model is
Planck and ACT-compatible. In Fig. \ref{plot2} we present the
confrontation of the model with Planck 2018 likelihood curves and
the ACT data for $\delta=[3003, 6003]$. As it can be seen, this
model too is well fitted within the data.

\subsection{Model III: Tensor-to-scalar ratio of the Form $r=\delta/N^6$}

Let us continue with the following scenario,
\begin{equation}
\label{r13} \centering r=\frac{\delta}{N^6} .
\end{equation}
\begin{figure}[h!]
\centering
\includegraphics[width=20pc]{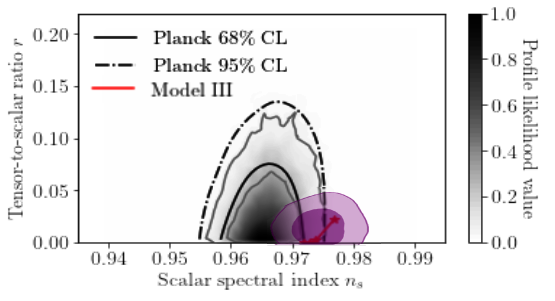}
\caption{Planck 2018 Likelihood Curves and ACT data for the Model
III with $r=\delta/N^6$ model}\label{plot3}
\end{figure}
From Eq. (\ref{difxi}) the scalar coupling function $\xi(N)$ is
obtained,
\begin{equation}
\label{xi13} \centering \xi(N)=\frac{1}{2} C_1
\left(\text{Ei}(N)-\frac{e^{N} (N+1)}{N^2}\right)+C_2 ,
\end{equation}
where $\text{Ei}(N)$ is the exponential integral function and
$C_1$, $C_2$ are dimensionless integration constants. The
corresponding potential can be found from (\ref{genpot}) and it
is,
\begin{equation}
\label{V13} \centering V(N)=\frac{3}{4 \kappa ^4 \left(\frac{1}{2}
C_1 \left(\text{Ei}(N)-\frac{e^{N} (N+1)}{N^2}\right)+C_2\right)}
\, .
\end{equation}
Accordingly, the function $\lambda(N)$ (\ref{functionE}) is,
\begin{equation}
\label{} \centering \lambda(N)=\frac{\delta  e^{-N} \left(C_1 N^2
\text{Ei}(N)-C_1 e^{N} (N+1)+2 C_2 N^2\right)}{16 C_1 N^5} \, ,
\end{equation}
and by using Eq. (\ref{tensorspectralindexfinal}) the tensor
spectral index becomes,
\begin{equation}
\label{nT13} \centering n_T=\frac{\delta  \left(-C_1 N^3
\text{Ei}(N)+C_1 e^{N} \left(N^2+N+1\right)-2 C_2 N^3\right)}{8
N^7 \left(C_1 N^2 \text{Ei}(N)-C_1 e^{N} (N+1)+2 C_2 N^2\right)} .
\end{equation}
The spectral index of the scalar perturbations is too complicated
to quote it here, so we refrain from quoting it here. The model at
hand is easily viable for a wide range of the parameter values.
For example, for $\delta=10^6$, $N=60$ and by choosing the free
parameters $C_1$ and $C_1$ as follows,
\begin{equation}\label{model1}
C_2=--5.570697\times 10^{20} C_1\, ,
\end{equation}
for $C_1=0.1249\times 10^{-4}$, we obtain
$n_{\mathcal{S}}=0.97412$, $r=0.021433$, $n_T=-0.516$ and finally
the predicted amplitude of the scalar perturbations is
$\mathcal{P}_{\zeta}(k_*)=2.15597\times 10^{-9}$, thus the model
is Planck and ACT-compatible. In Fig. \ref{plot3} we present the
confrontation of the model with Planck 2018 likelihood curves and
the ACT data for $\delta=[10^6, 10^9]$. As it can be seen, the
model is well fitted within the data.

\subsection{Model IV: Tensor-to-scalar ratio of the Form $r=\delta\,e^{-\beta N}$}

Finally, let us consider the following model,
\begin{equation}
\label{r14} \centering r=\delta\,e^{-\beta N}\, .
\end{equation}
\begin{figure}[h!]
\centering
\includegraphics[width=20pc]{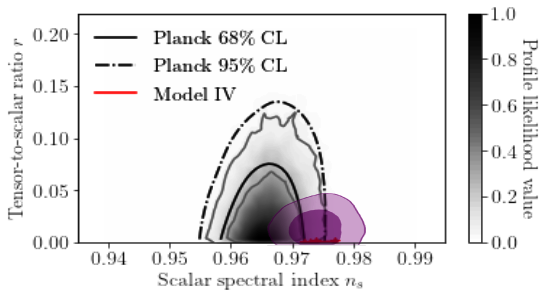}
\caption{Planck 2018 Likelihood Curves and ACT data for the Model
IV with $r=r=\delta\,e^{-\beta N}$ model}\label{plot4}
\end{figure}
From Eq. (\ref{difxi}) the scalar coupling function $\xi(N)$ is
obtained,
\begin{equation}
\label{xi1} \centering \xi(N)=C_2-\frac{2 C_1 e^{N-\frac{\beta
N}{2}}}{\beta -2} ,
\end{equation}
where $C_1$, $C_2$ are dimensionless integration constants. The
corresponding potential can be found from (\ref{genpot}) and it
is,
\begin{equation}
\label{V1} \centering V(N)=\frac{3}{4 \kappa ^4 \left(C_2-\frac{2
C_1 e^{N-\frac{\beta N}{2}}}{\beta -2}\right)} \, .
\end{equation}
Accordingly, the function $\lambda(N)$ (\ref{functionE}) is,
\begin{equation}
\label{} \centering \lambda(N)=\frac{1}{8} \delta  e^{\beta (-N)}
\left(\frac{C_2 e^{\frac{1}{2} (\beta -2) N}}{C_1}-\frac{2}{\beta
-2}\right) \, ,
\end{equation}
and by using Eq. (\ref{tensorspectralindexfinal}) the tensor
spectral index becomes,
\begin{equation}
\label{nT1} \centering n_T=\frac{\delta  \left((\beta +2) C_1
e^{N}-2 (\beta -2) C_2 e^{\frac{\beta N}{2}}\right)}{16 (\beta -2)
C_2 e^{\frac{3 \beta N}{2}}-32 C_1 e^{\beta N+N}} .
\end{equation}
The spectral index of the scalar perturbations is too complicated
to quote it here, so we refrain from quoting it here. The model at
hand is easily viable for a wide range of the parameter values.
For example, for $\delta=0.1$, $N=60$ and by choosing the free
parameters $C_1$ and $C_1$ as follows,
\begin{equation}\label{model1}
C_2=-5.9853\times 10^{24} C_1\, ,
\end{equation}
for $C_1=2\times 10^{-12}$, we obtain $n_{\mathcal{S}}=0.974$,
$r=0.00024787$, $n_T=-0.5166$ and finally the predicted amplitude
of the scalar perturbations is
$\mathcal{P}_{\zeta}(k_*)=2.15559\times 10^{-9}$, thus the model
is Planck and ACT-compatible. In Fig. \ref{plot4} we present the
confrontation of the model with Planck 2018 likelihood curves and
the ACT data for $\delta=[0.1, 0.4]$. As it can be seen, the model
is well fitted within the data.

\section{Conclusions}

In this article we considered GW170817-compatible
Einstein-Gauss-Bonnet inflationary theories which are compatible
with the ACT data using an inverse reconstruction technique.
Specifically we used a formalism that starts from a given form of
the tensor-to-scalar ratio, and the technique enables us to find
which Einstein-Gauss-Bonnet theory can yield such a
tensor-to-scalar ratio. Using the technique we found which
Gauss-Bonnet scalar coupling function and which potential can
yield the given tensor-to-scalar ratio. The technique specifically
allowed us to find the functions $\xi(N)$ and $V(N)$, so basically
functions of the $e$-foldings number. We also expressed the
derivatives of these functions, in terms of the $e$-foldings
number and we thus found analytical expressions for the slow-roll
indices and consequently for the spectral index of the scalar
perturbations and the tensor spectral index. We examined four
typical models which were simple to tackle analytically, and we
demonstrated that the resulting models yield ACT-compatible
phenomenology which is also compatible with the updated
BICEP/Planck constraints on the tensor-to-scalar ratio. The models
also yielded an amplitude of the scalar perturbations which was
also compatible with the Planck data. The models we presented
passed a series of truly severe constraints, starting from the
GW170817 constraints and proceeding to the compatibility with the
ACT and BICEP/Planck data. The reconstruction technique we used
here, can in principle be used in other frameworks, such as
$F(R,\phi)$ gravity, and we aim to extend this framework in such
theories.


\begin{thebibliography}{99}


\bibitem{TheLIGOScientific:2017qsa}
B.~P.~Abbott \textit{et al.} [LIGO Scientific and Virgo],
Phys. Rev. Lett. \textbf{119} (2017) no.16, 161101
doi:10.1103/PhysRevLett.119.161101 [arXiv:1710.05832 [gr-qc]].




\bibitem{Monitor:2017mdv}
B.~P.~Abbott \textit{et al.} [LIGO Scientific, Virgo, Fermi-GBM
and INTEGRAL],
Astrophys. J. Lett. \textbf{848} (2017) no.2, L13
doi:10.3847/2041-8213/aa920c [arXiv:1710.05834 [astro-ph.HE]].


\bibitem{GBM:2017lvd}
  B.~P.~Abbott {\it et al.}
  ``Multi-messenger Observations of a Binary Neutron Star Merger,''
  Astrophys.\ J.\  {\bf 848} (2017) no.2,  L12
  doi:10.3847/2041-8213/aa91c9
  [arXiv:1710.05833 [astro-ph.HE]].





\bibitem{Ezquiaga:2017ekz}
J.~M.~Ezquiaga and M.~Zumalac\'arregui,
Phys. Rev. Lett. \textbf{119} (2017) no.25, 251304
doi:10.1103/PhysRevLett.119.251304 [arXiv:1710.05901
[astro-ph.CO]].


\bibitem{Baker:2017hug}
T.~Baker, E.~Bellini, P.~G.~Ferreira, M.~Lagos, J.~Noller and
I.~Sawicki,
Phys. Rev. Lett. \textbf{119} (2017) no.25, 251301
doi:10.1103/PhysRevLett.119.251301 [arXiv:1710.06394
[astro-ph.CO]].


\bibitem{Creminelli:2017sry}
P.~Creminelli and F.~Vernizzi,
Phys. Rev. Lett. \textbf{119} (2017) no.25, 251302
doi:10.1103/PhysRevLett.119.251302 [arXiv:1710.05877
[astro-ph.CO]].


\bibitem{Sakstein:2017xjx}
J.~Sakstein and B.~Jain,
Phys. Rev. Lett. \textbf{119} (2017) no.25, 251303
doi:10.1103/PhysRevLett.119.251303 [arXiv:1710.05893
[astro-ph.CO]].

\bibitem{Fernandes:2022zrq}
P.~G.~S.~Fernandes, P.~Carrilho, T.~Clifton and D.~J.~Mulryne,
Class. Quant. Grav. \textbf{39} (2022) no.6, 063001
doi:10.1088/1361-6382/ac500a [arXiv:2202.13908 [gr-qc]].


\bibitem{Hwang:2005hb}
  J.~c.~Hwang and H.~Noh,
  Phys.\ Rev.\ D {\bf 71} (2005) 063536
  doi:10.1103/PhysRevD.71.063536
  [gr-qc/0412126].


\bibitem{Nojiri:2006je}
  S.~Nojiri, S.~D.~Odintsov and M.~Sami,
  Phys.\ Rev.\ D {\bf 74} (2006) 046004
  doi:10.1103/PhysRevD.74.046004
  [hep-th/0605039].




\bibitem{Cognola:2006sp}
  G.~Cognola, E.~Elizalde, S.~Nojiri, S.~Odintsov and S.~Zerbini,
  Phys.\ Rev.\ D {\bf 75} (2007) 086002
  doi:10.1103/PhysRevD.75.086002
  [hep-th/0611198].



\bibitem{Nojiri:2005vv}
  S.~Nojiri, S.~D.~Odintsov and M.~Sasaki,
  Phys.\ Rev.\ D {\bf 71} (2005) 123509
  doi:10.1103/PhysRevD.71.123509
  [hep-th/0504052].


\bibitem{Nojiri:2005jg}
  S.~Nojiri and S.~D.~Odintsov,
  Phys.\ Lett.\ B {\bf 631} (2005) 1
  doi:10.1016/j.physletb.2005.10.010
  [hep-th/0508049].







\bibitem{Satoh:2007gn}
  M.~Satoh, S.~Kanno and J.~Soda,
  Phys.\ Rev.\ D {\bf 77} (2008) 023526
  doi:10.1103/PhysRevD.77.023526
  [arXiv:0706.3585 [astro-ph]].



\bibitem{Bamba:2014zoa}
  K.~Bamba, A.~N.~Makarenko, A.~N.~Myagky and S.~D.~Odintsov,
  JCAP {\bf 1504} (2015) 001
  doi:10.1088/1475-7516/2015/04/001
  [arXiv:1411.3852 [hep-th]].


\bibitem{Yi:2018gse}
  Z.~Yi, Y.~Gong and M.~Sabir,
  Phys.\ Rev.\ D {\bf 98} (2018) no.8,  083521
  doi:10.1103/PhysRevD.98.083521
  [arXiv:1804.09116 [gr-qc]].


\bibitem{Guo:2009uk}
  Z.~K.~Guo and D.~J.~Schwarz,
  Phys.\ Rev.\ D {\bf 80} (2009) 063523
  doi:10.1103/PhysRevD.80.063523
  [arXiv:0907.0427 [hep-th]].


\bibitem{Guo:2010jr}
  Z.~K.~Guo and D.~J.~Schwarz,
  Phys.\ Rev.\ D {\bf 81} (2010) 123520
  doi:10.1103/PhysRevD.81.123520
  [arXiv:1001.1897 [hep-th]].


\bibitem{Jiang:2013gza}
  P.~X.~Jiang, J.~W.~Hu and Z.~K.~Guo,
  Phys.\ Rev.\ D {\bf 88} (2013) 123508
  doi:10.1103/PhysRevD.88.123508
  [arXiv:1310.5579 [hep-th]].




\bibitem{vandeBruck:2017voa}
  C.~van de Bruck, K.~Dimopoulos, C.~Longden and C.~Owen,
  arXiv:1707.06839 [astro-ph.CO].



\bibitem{Pozdeeva:2020apf}
E.~O.~Pozdeeva, M.~R.~Gangopadhyay, M.~Sami, A.~V.~Toporensky and
S.~Y.~Vernov,
Phys. Rev. D \textbf{102} (2020) no.4, 043525
doi:10.1103/PhysRevD.102.043525 [arXiv:2006.08027 [gr-qc]].



\bibitem{Vernov:2021hxo}
S.~Vernov and E.~Pozdeeva,
Universe \textbf{7} (2021) no.5, 149 doi:10.3390/universe7050149
[arXiv:2104.11111 [gr-qc]].


\bibitem{Pozdeeva:2021iwc}
E.~O.~Pozdeeva and S.~Y.~Vernov,
Eur. Phys. J. C \textbf{81} (2021) no.7, 633
doi:10.1140/epjc/s10052-021-09435-8 [arXiv:2104.04995 [gr-qc]].


\bibitem{Fomin:2020hfh}
I.~Fomin,
Eur. Phys. J. C \textbf{80} (2020) no.12, 1145
doi:10.1140/epjc/s10052-020-08718-w [arXiv:2004.08065 [gr-qc]].

\bibitem{DeLaurentis:2015fea}
  M.~De Laurentis, M.~Paolella and S.~Capozziello,
  Phys.\ Rev.\ D {\bf 91} (2015) no.8,  083531
  doi:10.1103/PhysRevD.91.083531
  [arXiv:1503.04659 [gr-qc]].


\bibitem{Chervon:2019sey}
   Scalar Field Cosmology, S.~Chervon, I.~Fomin, V.~Yurov and
   A.~Yurov, World Scientific 2019, \\  doi:10.1142/11405



\bibitem{Nozari:2017rta}
  K.~Nozari and N.~Rashidi,
  Phys.\ Rev.\ D {\bf 95} (2017) no.12,  123518
  doi:10.1103/PhysRevD.95.123518
  [arXiv:1705.02617 [astro-ph.CO]].




\bibitem{Odintsov:2018zhw}
  S.~D.~Odintsov and V.~K.~Oikonomou,
  Phys.\ Rev.\ D {\bf 98} (2018) no.4,  044039
  doi:10.1103/PhysRevD.98.044039
  [arXiv:1808.05045 [gr-qc]].


  \bibitem{Kawai:1998ab}
  S.~Kawai, M.~a.~Sakagami and J.~Soda,
  Phys.\ Lett.\ B {\bf 437}, 284 (1998)
  doi:10.1016/S0370-2693(98)00925-3
  [gr-qc/9802033].


\bibitem{Yi:2018dhl}
  Z.~Yi and Y.~Gong,
  Universe {\bf 5} (2019) no.9,  200
  doi:10.3390/universe5090200
  [arXiv:1811.01625 [gr-qc]].


\bibitem{vandeBruck:2016xvt}
  C.~van de Bruck, K.~Dimopoulos and C.~Longden,
  Phys.\ Rev.\ D {\bf 94} (2016) no.2,  023506
  doi:10.1103/PhysRevD.94.023506
  [arXiv:1605.06350 [astro-ph.CO]].




\bibitem{Maeda:2011zn}
  K.~i.~Maeda, N.~Ohta and R.~Wakebe,
  Eur.\ Phys.\ J.\ C {\bf 72} (2012) 1949
  doi:10.1140/epjc/s10052-012-1949-6
  [arXiv:1111.3251 [hep-th]].


\bibitem{Ai:2020peo}
W.~Y.~Ai,
Commun. Theor. Phys. \textbf{72} (2020) no.9, 095402
doi:10.1088/1572-9494/aba242 [arXiv:2004.02858 [gr-qc]].





\bibitem{Easther:1996yd}
  R.~Easther and K.~i.~Maeda,
  Phys.\ Rev.\ D {\bf 54} (1996) 7252
  doi:10.1103/PhysRevD.54.7252
  [hep-th/9605173].



\bibitem{Codello:2015mba}
A.~Codello and R.~K.~Jain,
Class. Quant. Grav. \textbf{33} (2016) no.22, 225006
doi:10.1088/0264-9381/33/22/225006 [arXiv:1507.06308 [gr-qc]].






\bibitem{Oikonomou:2021kql}
V.~K.~Oikonomou,
Class. Quant. Grav. \textbf{38} (2021) no.19, 195025
doi:10.1088/1361-6382/ac2168 [arXiv:2108.10460 [gr-qc]].



\bibitem{Oikonomou:2022xoq}
V.~K.~Oikonomou,
Astropart. Phys. \textbf{141} (2022), 102718
doi:10.1016/j.astropartphys.2022.102718 [arXiv:2204.06304
[gr-qc]].


\bibitem{Odintsov:2020sqy}
S.~D.~Odintsov, V.~K.~Oikonomou and F.~P.~Fronimos,
Nucl. Phys. B \textbf{958} (2020), 115135
doi:10.1016/j.nuclphysb.2020.115135 [arXiv:2003.13724 [gr-qc]].



\bibitem{Oikonomou:2024etl}
V.~K.~Oikonomou,
Phys. Lett. B \textbf{856} (2024), 138890
doi:10.1016/j.physletb.2024.138890 [arXiv:2407.12155 [gr-qc]].

\bibitem{Fier:2025huc}
J.~Fier, H.~Han, B.~Li, K.~Lin, S.~Mukohyama and A.~Wang,
[arXiv:2503.01975 [gr-qc]].



\bibitem{Oikonomou:2022ksx}
V.~K.~Oikonomou, P.~D.~Katzanis and I.~C.~Papadimitriou,
Class. Quant. Grav. \textbf{39} (2022) no.9, 095008
doi:10.1088/1361-6382/ac5eba [arXiv:2203.09867 [gr-qc]].



\bibitem{inflation1}
 A.~D.~Linde,
 Lect.\ Notes Phys.\ {\bf 738} (2008) 1
 [arXiv:0705.0164 [hep-th]].

\bibitem{inflation2} D.~S.~Gorbunov and V.~A.~Rubakov,
``Introduction to the theory of the early universe: Cosmological
perturbations and inflationary theory,'' Hackensack, USA: World
Scientific (2011) 489 p;
%


\bibitem{inflation3}A.~Linde,
arXiv:1402.0526 [hep-th];


\bibitem{inflation4}D.~H.~Lyth and A.~Riotto,
Phys.\ Rept.\  {\bf 314} (1999) 1 [hep-ph/9807278].



\bibitem{SimonsObservatory:2019qwx}
M.~H.~Abitbol \textit{et al.} [Simons Observatory],
Bull. Am. Astron. Soc. \textbf{51} (2019), 147 [arXiv:1907.08284
[astro-ph.IM]].



\bibitem{LiteBIRD:2022cnt}
E.~Allys \textit{et al.} [LiteBIRD],
PTEP \textbf{2023} (2023) no.4, 042F01 doi:10.1093/ptep/ptac150
[arXiv:2202.02773 [astro-ph.IM]].




\bibitem{Hild:2010id}
S.~Hild, M.~Abernathy, F.~Acernese, P.~Amaro-Seoane, N.~Andersson,
K.~Arun, F.~Barone, B.~Barr, M.~Barsuglia and M.~Beker, \textit{et
al.}
Class. Quant. Grav. \textbf{28} (2011), 094013
doi:10.1088/0264-9381/28/9/094013 [arXiv:1012.0908 [gr-qc]].




\bibitem{Baker:2019nia}
J.~Baker, J.~Bellovary, P.~L.~Bender, E.~Berti, R.~Caldwell,
J.~Camp, J.~W.~Conklin, N.~Cornish, C.~Cutler and R.~DeRosa,
\textit{et al.}
[arXiv:1907.06482 [astro-ph.IM]].


\bibitem{Smith:2019wny}
T.~L.~Smith and R.~Caldwell,
Phys. Rev. D \textbf{100} (2019) no.10, 104055
doi:10.1103/PhysRevD.100.104055 [arXiv:1908.00546 [astro-ph.CO]].


\bibitem{Crowder:2005nr}
J.~Crowder and N.~J.~Cornish,
Phys. Rev. D \textbf{72} (2005), 083005
doi:10.1103/PhysRevD.72.083005 [arXiv:gr-qc/0506015 [gr-qc]].


\bibitem{Smith:2016jqs}
T.~L.~Smith and R.~Caldwell,
Phys. Rev. D \textbf{95} (2017) no.4, 044036
doi:10.1103/PhysRevD.95.044036 [arXiv:1609.05901 [gr-qc]].



\bibitem{Seto:2001qf}
N.~Seto, S.~Kawamura and T.~Nakamura,
Phys. Rev. Lett. \textbf{87} (2001), 221103
doi:10.1103/PhysRevLett.87.221103 [arXiv:astro-ph/0108011
[astro-ph]].


\bibitem{Kawamura:2020pcg}
S.~Kawamura, M.~Ando, N.~Seto, S.~Sato, M.~Musha, I.~Kawano,
J.~Yokoyama, T.~Tanaka, K.~Ioka and T.~Akutsu, \textit{et al.}
[arXiv:2006.13545 [gr-qc]].



\bibitem{Bull:2018lat}
A.~Weltman, P.~Bull, S.~Camera, K.~Kelley, H.~Padmanabhan,
J.~Pritchard, A.~Raccanelli, S.~Riemer-S\o{}rensen, L.~Shao and
S.~Andrianomena, \textit{et al.}
Publ. Astron. Soc. Austral. \textbf{37} (2020), e002
doi:10.1017/pasa.2019.42 [arXiv:1810.02680 [astro-ph.CO]].




\bibitem{LISACosmologyWorkingGroup:2022jok}
P.~Auclair \textit{et al.} [LISA Cosmology Working Group],
[arXiv:2204.05434 [astro-ph.CO]].


\bibitem{NANOGrav:2023gor}
G.~Agazie \textit{et al.} [NANOGrav],
Astrophys. J. Lett. \textbf{951} (2023) no.1, L8
doi:10.3847/2041-8213/acdac6 [arXiv:2306.16213 [astro-ph.HE]].



\bibitem{Vagnozzi:2023lwo}
S.~Vagnozzi,
JHEAp \textbf{39} (2023), 81-98 doi:10.1016/j.jheap.2023.07.001
[arXiv:2306.16912 [astro-ph.CO]].


\bibitem{Oikonomou:2023qfz}
V.~K.~Oikonomou,
Phys. Rev. D \textbf{108} (2023) no.4, 043516
doi:10.1103/PhysRevD.108.043516 [arXiv:2306.17351 [astro-ph.CO]].




\bibitem{reviews1}
 S.~Nojiri, S.~D.~Odintsov and V.~K.~Oikonomou,
  Phys.\ Rept.\  {\bf 692} (2017) 1
  [arXiv:1705.11098 [gr-qc]].

\bibitem{reviews2}


 S. Capozziello, M. De Laurentis,
   Phys.\ Rept.\  {\bf 509}, 167 (2011);\\
 V.~Faraoni and S.~Capozziello,
  Fundam.\ Theor.\ Phys.\  {\bf 170} (2010).



\bibitem{reviews3}
S. Nojiri, S.D. Odintsov,
  eConf {\bf C0602061}, 06 (2006)
  [Int.\ J.\ Geom.\ Meth.\ Mod.\ Phys.\  {\bf 4}, 115 (2007)].


   \bibitem{reviews4}

S. Nojiri, S.D. Odintsov,
   Phys.\ Rept.\  {\bf 505}, 59 (2011);





\bibitem{ACT:2025fju}
T.~Louis \textit{et al.} [ACT],
[arXiv:2503.14452 [astro-ph.CO]].


\bibitem{ACT:2025tim}
E.~Calabrese \textit{et al.} [ACT],
[arXiv:2503.14454 [astro-ph.CO]].


\bibitem{BICEP:2021xfz}
P.~A.~R.~Ade \textit{et al.} [BICEP and Keck],
Phys. Rev. Lett. \textbf{127} (2021) no.15, 151301
doi:10.1103/PhysRevLett.127.151301 [arXiv:2110.00483
[astro-ph.CO]].



\bibitem{Kallosh:2025rni}
R.~Kallosh, A.~Linde and D.~Roest,
[arXiv:2503.21030 [hep-th]].


\bibitem{Gao:2025onc}
Q.~Gao, Y.~Gong, Z.~Yi and F.~Zhang,
[arXiv:2504.15218 [astro-ph.CO]].



\bibitem{Liu:2025qca}
L.~Liu, Z.~Yi and Y.~Gong,
[arXiv:2505.02407 [astro-ph.CO]].

\bibitem{Yogesh:2025wak}
Yogesh, A.~Mohammadi, Q.~Wu and T.~Zhu,
[arXiv:2505.05363 [astro-ph.CO]].

\bibitem{Yi:2025dms}
Z.~Yi, X.~Wang, Q.~Gao and Y.~Gong,
[arXiv:2505.10268 [astro-ph.CO]].


\bibitem{Peng:2025bws}
Z.~Z.~Peng, Z.~C.~Chen and L.~Liu,
[arXiv:2505.12816 [astro-ph.CO]].

\bibitem{Yin:2025rrs}
W.~Yin,
[arXiv:2505.03004 [hep-ph]].

\bibitem{Byrnes:2025kit}
C.~T.~Byrnes, M.~Cort\^es and A.~R.~Liddle,
[arXiv:2505.09682 [astro-ph.CO]].

\bibitem{Wolf:2025ecy}
W.~J.~Wolf,
[arXiv:2506.12436 [astro-ph.CO]].

\bibitem{Aoki:2025wld}
S.~Aoki, H.~Otsuka and R.~Yanagita,
[arXiv:2504.01622 [hep-ph]].


\bibitem{Gao:2025viy}
Q.~Gao, Y.~Qian, Y.~Gong and Z.~Yi,
[arXiv:2506.18456 [gr-qc]].



\bibitem{Zahoor:2025nuq}
M.~Zahoor, S.~Khan and I.~A.~Bhat,
[arXiv:2507.18684 [astro-ph.CO]].


\bibitem{Ferreira:2025lrd}
E.~G.~M.~Ferreira, E.~McDonough, L.~Balkenhol, R.~Kallosh, L.~Knox
and A.~Linde,
[arXiv:2507.12459 [astro-ph.CO]].


\bibitem{Mohammadi:2025gbu}
A.~Mohammadi, Yogesh and A.~Wang,
[arXiv:2507.06544 [astro-ph.CO]].

\bibitem{Choudhury:2025vso}
S.~Choudhury, G.~Bauyrzhan, S.~K.~Singh and K.~Yerzhanov,
[arXiv:2506.15407 [astro-ph.CO]].

\bibitem{Odintsov:2025wai}
S.~D.~Odintsov and V.~K.~Oikonomou,
Phys. Lett. B \textbf{868} (2025), 139779
doi:10.1016/j.physletb.2025.139779 [arXiv:2506.08193 [gr-qc]].


\bibitem{Q:2025ycf}
R.~D.~A.~Q., J.~Chagoya and A.~A.~Roque,
[arXiv:2508.13273 [gr-qc]].


\bibitem{Zhu:2025twm}
Y.~Zhu, Q.~Gao, Y.~Gong and Z.~Yi,
[arXiv:2508.09707 [astro-ph.CO]].

\bibitem{Kouniatalis:2025orn}
G.~Kouniatalis and E.~N.~Saridakis,
[arXiv:2507.17721 [astro-ph.CO]].


\bibitem{Hai:2025wvs}
M.~Hai, A.~R.~Kamal, N.~F.~Shamma and M.~S.~J.~Shuvo,
[arXiv:2506.08083 [hep-th]].


\bibitem{Dioguardi:2025vci}
C.~Dioguardi, A.~J.~Iovino and A.~Racioppi,
Phys. Lett. B \textbf{868} (2025), 139664
doi:10.1016/j.physletb.2025.139664 [arXiv:2504.02809 [gr-qc]].

\bibitem{Yuennan:2025kde}
J.~Yuennan, P.~Koad, F.~Atamurotov and P.~Channuie,
[arXiv:2508.17263 [astro-ph.CO]].


\bibitem{Kuralkar:2025zxr}
H.~J.~Kuralkar, S.~Panda and A.~Vidyarthi,
[arXiv:2504.15061 [gr-qc]].


\bibitem{Kuralkar:2025hoz}
H.~J.~Kuralkar, S.~Panda and A.~Vidyarthi,
JCAP \textbf{05} (2025), 073 doi:10.1088/1475-7516/2025/05/073
[arXiv:2502.03573 [astro-ph.CO]].

\bibitem{Modak:2025bjv}
T.~Modak,
[arXiv:2509.02979 [astro-ph.CO]].

\bibitem{Oikonomou:2025xms}
V.~K.~Oikonomou,
[arXiv:2508.19196 [gr-qc]].



\bibitem{Odintsov:2025jky}
S.~D.~Odintsov and V.~K.~Oikonomou,
[arXiv:2508.17358 [gr-qc]].


\bibitem{Aoki:2025ywt}
S.~Aoki, H.~Otsuka and R.~Yanagita,
[arXiv:2509.06739 [hep-ph]].






\bibitem{Ahghari:2025hfy}
Z.~Ahghari and M.~Farhoudi,
[arXiv:2512.12286 [gr-qc]].


\bibitem{McDonough:2025lzo}
E.~McDonough and E.~G.~M.~Ferreira,
[arXiv:2512.05108 [astro-ph.CO]].



\bibitem{Chakraborty:2025wqn}
D.~Chakraborty, M.~Hai, S.~T.~Jahan, A.~R.~Kamal and
M.~S.~J.~Shuvo,
[arXiv:2511.19610 [hep-th]].



\bibitem{NooriGashti:2025gug}
S.~Noori Gashti, M.~A.~S.~Afshar, M.~R.~Alipour, B.~Pourhassan and
J.~Sadeghi,
Eur. Phys. J. C \textbf{85} (2025) no.11, 1343
doi:10.1140/epjc/s10052-025-15066-0

\bibitem{Yuennan:2025mlg}
J.~Yuennan, F.~Atamurotov and P.~Channuie,
[arXiv:2511.17216 [astro-ph.CO]].






\bibitem{Deb:2025gtk}
B.~Deb and A.~Deshamukhya,
[arXiv:2511.06453 [gr-qc]].



\bibitem{Afshar:2025ndm}
M.~A.~S.~Afshar, S.~Noori Gashti, M.~R.~Alipour, B.~Pourhassan,
I.~Sakalli and J.~Sadeghi,
[arXiv:2510.20876 [astro-ph.CO]].

\bibitem{Ellis:2025zrf}
J.~Ellis, M.~A.~G.~Garcia, K.~A.~Olive and S.~Verner,
[arXiv:2510.18656 [hep-ph]].


\bibitem{Yuennan:2025tyx}
J.~Yuennan, F.~Atamurotov and P.~Channuie,
Phys. Lett. B \textbf{872} (2026), 140065
doi:10.1016/j.physletb.2025.140065 [arXiv:2509.23329 [gr-qc]].


\bibitem{Wang:2025cpp}
Q.~Y.~Wang,
[arXiv:2512.10862 [astro-ph.CO]].


\bibitem{Qiu:2025uot}
Z.~C.~Qiu, Y.~H.~Pang and Q.~G.~Huang,
[arXiv:2510.18320 [astro-ph.CO]].

\bibitem{Wang:2025dbj}
X.~Wang, K.~Kohri and T.~T.~Yanagida,
[arXiv:2506.06797 [astro-ph.CO]].

\bibitem{Asaka:2015vza}
T.~Asaka, S.~Iso, H.~Kawai, K.~Kohri, T.~Noumi and T.~Terada,
PTEP \textbf{2016} (2016) no.12, 123E01 doi:10.1093/ptep/ptw161
[arXiv:1507.04344 [hep-th]].


\bibitem{Oikonomou:2025htz}
V.~K.~Oikonomou,
Phys. Lett. B \textbf{871} (2025), 139972
doi:10.1016/j.physletb.2025.139972 [arXiv:2508.17363 [gr-qc]].

\bibitem{Choudhury:2025hnu}
S.~Choudhury, S.~K.~Singh and S.~K.~Sahoo,
[arXiv:2511.19898 [gr-qc]].


\bibitem{Singh:2025uyr}
S.~K.~Singh,
[arXiv:2511.05545 [hep-ph]].



\bibitem{Kim:2025dyi}
J.~Kim, X.~Wang, Y.~l.~Zhang and Z.~Ren,
JCAP \textbf{09} (2025), 011 doi:10.1088/1475-7516/2025/09/011
[arXiv:2504.12035 [astro-ph.CO]].

\bibitem{Peng:2026ofs}
Z.~Y.~Peng, H.~S.~Yuan, Q.~Lai, J.~Q.~Jiang, G.~Ye, J.~Zhang and
Y.~S.~Piao,
[arXiv:2601.14288 [astro-ph.CO]].



\bibitem{Planck:2018jri}
Y.~Akrami \textit{et al.} [Planck],
Astron. Astrophys. \textbf{641} (2020), A10
doi:10.1051/0004-6361/201833887 [arXiv:1807.06211 [astro-ph.CO]].

\end{thebibliography}
\end{document}